\documentclass[twocolumn,superscriptaddress,aps,preprintnumbers,amsmath,amssymb,prl,nofootinbib]{revtex4}

\usepackage{graphicx}
\usepackage{epstopdf}
\usepackage{dcolumn}
\usepackage{bm}
\usepackage{hyperref}
\usepackage{color}

\begin{document}

\def\a{\alpha}
\def\b{\beta}
\def\c{\varepsilon}
\def\d{\delta}
\def\e{\epsilon}
\def\f{\phi}
\def\g{\gamma}
\def\h{\theta}
\def\k{\kappa}
\def\l{\lambda}
\def\m{\mu}
\def\n{\nu}
\def\p{\psi}
\def\q{\partial}
\def\r{\rho}
\def\s{\sigma}
\def\t{\tau}
\def\u{\upsilon}
\def\v{\varphi}
\def\w{\omega}
\def\x{\xi}
\def\y{\eta}
\def\z{\zeta}
\def\D{\Delta}
\def\G{\Gamma}
\def\H{\Theta}
\def\L{\Lambda}
\def\F{\Phi}
\def\P{\Psi}
\def\S{\Sigma}

\def\o{\over}
\def\beq{\begin{eqnarray}}
\def\eeq{\end{eqnarray}}
\newcommand{\gsim}{ \mathop{}_{\textstyle \sim}^{\textstyle >} }
\newcommand{\lsim}{ \mathop{}_{\textstyle \sim}^{\textstyle <} }
\newcommand{\vev}[1]{ \left\langle {#1} \right\rangle }
\newcommand{\bra}[1]{ \langle {#1} | }
\newcommand{\ket}[1]{ | {#1} \rangle }
\newcommand{\EV}{ {\rm eV} }
\newcommand{\KEV}{ {\rm keV} }
\newcommand{\MEV}{ {\rm MeV} }
\newcommand{\GEV}{ {\rm GeV} }
\newcommand{\TEV}{ {\rm TeV} }
\newcommand{\1}{\mbox{1}\hspace{-0.25em}\mbox{l}}
\def\diag{\mathop{\rm diag}\nolimits}
\def\Spin{\mathop{\rm Spin}}
\def\SO{\mathop{\rm SO}}
\def\O{\mathop{\rm O}}
\def\SU{\mathop{\rm SU}}
\def\U{\mathop{\rm U}}
\def\Sp{\mathop{\rm Sp}}
\def\SL{\mathop{\rm SL}}
\def\tr{\mathop{\rm tr}}

\def\IJMP{Int.~J.~Mod.~Phys. }
\def\MPL{Mod.~Phys.~Lett. }
\def\NP{Nucl.~Phys. }
\def\PL{Phys.~Lett. }
\def\PR{Phys.~Rev. }
\def\PRL{Phys.~Rev.~Lett. }
\def\PTP{Prog.~Theor.~Phys. }
\def\ZP{Z.~Phys. }


\title{Dynamical Chaotic Inflation in the Light of BICEP2}

\author{Keisuke Harigaya}
\affiliation{Kavli IPMU (WPI), TODIAS, University of Tokyo, Kashiwa, 277-8583, Japan}
\author{Masahiro Ibe}
\affiliation{Kavli IPMU (WPI), TODIAS, University of Tokyo, Kashiwa, 277-8583, Japan}
\affiliation{ICRR, University of Tokyo, Kashiwa, 277-8582, Japan}
\author{Kai Schmitz}
\affiliation{Kavli IPMU (WPI), TODIAS, University of Tokyo, Kashiwa, 277-8583, Japan}
\author{Tsutomu T.~Yanagida}
\affiliation{Kavli IPMU (WPI), TODIAS, University of Tokyo, Kashiwa, 277-8583, Japan}

\begin{abstract}
The measurement of a large tensor-to-scalar ratio by the BICEP2 experiment,
$r = 0.20_{-0.05}^{+0.07}$, 
severely restricts the landscape of viable inflationary models and shifts
attention once more towards models featuring large inflaton field values.
In this context, chaotic inflation based on a fractional power-law potential
that is dynamically generated by the dynamics of a strongly coupled supersymmetric gauge theory
appears to be particularly attractive.
We revisit this class of inflation models and find that,
in the light of the BICEP2 measurement, models with a non-minimal gauge group behind 
the dynamical model seem to be disfavored, while the model with the simplest group,
i.e.\ $SU(2)$, is consistent with all results.
We also discuss how the dynamical model can be distinguished from
the standard chaotic inflation model based on a quadratic inflaton potential.
\end{abstract}

\date{\today}
\maketitle
\preprint{IPMU 14-0061}

\section{Introduction}
Cosmic inflation\,\cite{Guth:1980zm} is an enormously successful paradigm 
of modern cosmology, which not only explains why the 
universe is almost homogeneous and spatially flat but which also accounts for the
origin of the anisotropies in the Cosmic Microwave Background (CMB) radiation
as well as for the origin of the Large Scale Structure of the Universe\,\cite{Mukhanov:1981xt,Lyth:1998xn}.
Among the various models of inflation, chaotic inflation\,\cite{Linde:1983gd} 
is particularly attractive since it is free from the initial condition
problem at the Planck time.
Moreover, the large field values typically encountered in models of chaotic inflation
predict a large contribution from gravitational waves to the CMB power spectrum\,\cite{Starobinsky:1985ww},
which can be tested by measuring the so-called B-mode of the CMB polarization spectrum.
Interestingly, the BICEP2 collaboration recently announced the first measurement
of just such a B-mode signal, corresponding 
to a tensor-to-scalar ratio of $r=0.20^{+0.07}_{-0.05}$ at $1\,\sigma$\,\cite{BICEP2},
which strongly suggests the presence of primordial B-mode polarization in the CMB.%
\footnote{As pointed out in\,\cite{BICEP2}, the observed ratio,
$r=0.20^{+0.07}_{-0.05}$, is in tension with the upper limit on this ratio,
$r<0.11$ (at 95\%C.L.)\cite{Ade:2013zuv}, which is
deduced from a combination of Planck, SPT and ACT data with
polarization data from WMAP.
In the following discussion, we shall keep this tension in mind
when applying the BICEP2 result to our dynamical model
of chaotic inflation.\smallskip}

This recent progress motivates us to revisit
\textit{dynamical chaotic inflation}, which was proposed
in\,\cite{Harigaya:2012pg} and in which  the inflaton potential is
generated by the dynamics of a simple strongly coupled supersymmetric gauge theory.
One prominent feature of this class of models is that it
predicts a fractional power-law potential for the inflaton
with the fractional power being $1$ or smaller.%
\footnote{Such potentials can also be realized by
introducing a running kinetic term for the inflaton\,\cite{Takahashi:2010ky}.
In string theory, fractional power-law potentials have been
derived in \cite{Silverstein:2008sg}. For dynamical chaotic inflation featuring
fractional powers larger than 1, cf.\,\cite{Harigaya:2014zz}.}
Chaotic inflation of this type can be distinguished from the simplest versions of chaotic inflation, 
i.e.\ models with a quadratic or quartic potential,
by precise measurements of the inflationary CMB observables.
Furthermore, dynamical chaotic inflation is also attractive
since it entails that the energy scale of inflation 
is generated via dimensional transmutation due to the strong gauge dynamics.
This provides an explanation for why inflation takes place at energies much below the Planck scale.%
\footnote{For other examples of models in which the scale of inflation
is generated dynamically, cf.\ Refs.\,\cite{Dimopoulos:1997fv}.\smallskip}

The organization of the paper is as follows.
First, we review  chaotic inflation 
emerging from a strongly coupled supersymmetric gauge theory,
which eventually leads us to an inflaton potential proportional
to some fractional power of the inflaton field.
Then, we discuss how the value of the tensor-to-scalar ratio observed 
by BICEP2 can be explained in this class of models.

\section{Dynamical Generation of the Inflaton Potential}
\label{sec:potential}
Let us briefly review the scenario of dynamical chaotic inflation,
in which strong gauge interactions such as those
proposed in\,\cite{Harigaya:2012pg} are responsible for the
dynamical generation of the inflaton potential.
For that purpose, let us consider an $SP(N)$ supersymmetric gauge theory%
\footnote{In our convention $SP(1)$ is equivalent to $SU(2)$.
Alternative strong gauge groups, such as $SU(N)$, will
be considered in \cite{Harigaya:2014zz}.}
with $2(N+2)$  chiral superfields in the fundamental representation,
$Q^I$ ($I=1\cdots 2(N+2)$).
Besides these fundamental representations, we also introduce
$(N+2)(2N+3)$ gauge-singlet chiral superfields $\mathcal{Z}_{IJ}$ ($=-\mathcal{Z}_{JI}$),
which couple to the fundamental representations in the superpotential via
\begin{eqnarray}
\label{eq:tree1}
 W = \frac{1}{2}\l_{IJ} \mathcal{Z}_{IJ}Q^IQ^J\ ,
\end{eqnarray}
with coupling constants $\l_{IJ}$, which we shall assume to
be close to each other in the following,
i.e.\ $\lambda_{IJ} \simeq \lambda$, for simplicity.
Note that the couplings to the gauge singlets $\mathcal{Z}_{IJ}$
in Eq.~\eqref{eq:tree1} lifts all of the quantum moduli, $Q^IQ^J$.

To see how the inflaton potential is generated, imagine that one of the singlet fields, 
$S = \mathcal{Z}_{(2N+3)(2N+4)}$ for instance, has a very large field value, so that 
the effective mass of $Q^{I= 2N+3}$ and $Q^{J=2N+4}$ becomes
much larger than the dynamical scale of the $SP(N)$
gauge interactions $\L$.
Then, $Q^{I= 2N+3}$ and $Q^{J=2N+4}$ 
decouple perturbatively and the model reduces to an $SP(N)$ gauge theory 
with $2(N+1)$ fundamentals and $(N+1)(2N+1)$ singlets.
This low energy effective theory is nothing but the dynamical supersymmetry
breaking model proposed in\,\cite{Izawa:1996pk}.
Therefore, for a given non-vanishing $S$,
the model breaks supersymmetry dynamically and results
in a ``vacuum energy" that depends on the field value of $S$,
\begin{eqnarray}
\label{eq:spot2}
V \simeq \lambda^2 (N+1) \Lambda_{\rm eff}^4(S) \simeq
\l^2 (N+1)\L^4    \left(\frac{\l\left|S\right|}{\Lambda}\right)^{\frac{2}{N+1}}\ .
\end{eqnarray}
where we have substituted the effective dynamical scale
\begin{eqnarray}
 \L_{\rm eff} = \L \times\left( \frac{\l\left|S\right|}{\L} \right)^{\frac{1}{2(N+1)}}\ ,
\end{eqnarray}
for field values $\lambda S \gg \Lambda$.

As a result, we find that the scalar component of the singlet $S$
obtains a fractional power-law potential,
\begin{eqnarray}
 V \propto \left|S\right|^{p}\ ,
\end{eqnarray}
in which the power is solely determined by the size of the $SP(N)$ gauge group,%
\footnote{In this letter, we only discuss the scalar potential for large field values,
$\lambda S \gg \Lambda$.
The vacuum structure for $\lambda S \ll \Lambda$ has been addressed
in\,\cite{Harigaya:2012pg}.
}
\begin{eqnarray}
 p = \frac{2}{N+1}\ .
 \label{eq:fracpower}
\end{eqnarray} 
For example, for $SU(2) = SP(1)$, we obtain a linear potential,
while a much flatter potential is generated for $N\gg 1$.
In everything what follows, we will assume that the field $S$ plays the role of the
inflaton, although any of the other singlets could be equally used as well.
Finally, we remark that it is easy to to generalize dynamical chaotic inflation
and in particular Eq.~\eqref{eq:fracpower}, such that $p$ can also
take fractional values larger than $1$, cf.\,\cite{Harigaya:2014zz}.

During chaotic inflation, the field value of the inflaton
exceeds the Planck scale $M_{\rm Pl}$.
Before we can be sure that the above model allows for a successful
implementation of chaotic inflation, we thus
have to carefully examine the supergravity contributions 
to the scalar potential.
For example, naively coupling the above model to supergravity by simply assuming
a minimal K\"ahler potential, $K=S^\dagger S$, leads to a very steep scalar potential
\begin{eqnarray}
 V \simeq e^{\left|S\right|^2/M_{\rm Pl}^2}\times  \l^2 (N+1)\L^4 
 \left(\frac{\l\left|S\right|}{\Lambda}\right)^{\frac{2}{N+1}}\ ,
\end{eqnarray}
above the Planck scale.
To avoid such a steep potential, we assume a shift symmetry
in the direction of $S$\,\cite{Kawasaki:2000yn,Kallosh:2011qk},
\begin{eqnarray}
\label{eq:shift}
 S \to S + i c\ ,\quad c\in {\mathbb R} \ ,
\end{eqnarray}
(cf.\ also\,\cite{Kallosh:2010ug} for recent developments)
which renders the K\"ahler potential a function of $S+S^\dagger$ only, 
\begin{eqnarray}
\label{eq:Kshif}
 K  = \frac{1}{2}\left( S + S^\dagger \right)^2 + \cdots\ ,
\end{eqnarray}
such that it no longer depends on $\Im(S)$, the imaginary component of $S$.
Consequently, the imaginary component of $S$ merely has a fractional power-law potential
even for $\Im(S) \gg M_{\rm Pl}$.
In the following,  $\phi=\sqrt{2}\Im(S)$ is identified as the inflaton
in the scenario of chaotic inflation based on the dynamically generated fractional power-law potential 
in Eq.\,(\ref{eq:spot2}).

It should be noted that the shift symmetry introduced in Eq.~\eqref{eq:shift}
is explicitly broken by the Yukawa interactions in Eq.\,(\ref{eq:tree1}), which 
induce the K\"ahler potential
\begin{eqnarray}
\delta K \sim \frac{2N\l^2}{16\pi^2} |S|^2 \log \left(\frac{\mu^2}{M_{\rm Pl}^2}\right)\ ,
\end{eqnarray}
where $\m$ is a renormalization scale.%
\footnote{Here, we have assumed that the shift-symmetric K\"ahler potential
in Eq.\,(\ref{eq:Kshif}) is defined around the Planck scale.\smallskip}
This breaking term leads again to a steep exponential potential 
for $\Im(S)$ once $\Im(S)\gg M_{\rm Pl}$.
To avoid such dangerous effects, we therefore assume
that $\l$ is  rather suppressed, $\lambda\ll O( 10^{-1})$.%
\footnote{Small $\lambda$ is also required in order to keep the effective $Q$ mass
below the Planck scale even during inflation, $S \sim 10\cdots100\,M_{\textrm{Pl}}$,
i.e.\ $\lambda S \ll M_{\rm Pl}$.}


\section{Testing dynamical chaotic inflation}
As we have demonstrated, simple strongly coupled gauge dynamics are able to
generate an inflationary potential featuring a fractional power.
In this section, we now outline how chaotic inflation proceeds in these models.
We also summarize the predictions for the inflationary observables encoded in
the CMB power spectrum.

Inflation starts out at an arbitrary initial value of the inflaton field $S = i\phi/\sqrt{2}$
above the Planck scale, $\phi \gg M_{\textrm{Pl}}$.
There, the $SP(N)$ gauge interactions are perturbative and 
inflation is characterized by the slow-roll motion of the inflaton in the
effective potential in Eq.\,(\ref{eq:spot2}) towards smaller field values.
We assume that, during inflation, the strong gauge dynamics are negligible,
which requires $\lambda p M_{\rm Pl} \gg \Lambda$.
Inflation finally ends once the slow-roll conditions are no longer
satisfied, which happens when the inflaton field reaches $\phi \simeq pM_{\rm Pl}$.
Well after inflation, the inflaton oscillates around its origin 
with a mass of $m_\phi \simeq \lambda \Lambda$.
At small field values, the strongly interacting theory is in
the $s$-confinement phase\,\cite{Seiberg:1994bz,Intriligator:1995ne},
which is well-behaved and free of singularities at the origin in field space.%
\footnote{At intermediate field values, $\lambda \phi \simeq \Lambda$, 
where the gauge dynamics transit from the perturbative to the strongly coupled picture,
we lack the ability to precisely calculate the inflaton potential.
In our discussion, we assume that the effective inflaton potential
exhibits no peculiar features around $\Lambda / \lambda$, but that it is
instead smoothly connected from one regime to the other.\smallskip}

After inflation, the inflaton finally decays into radiation through appropriate operators.
For example, the reheating temperature can be estimated as
\begin{eqnarray}
\label{eq:TRH1}
T_{R,{\rm dim} 5} \sim 10^{12-13}{\rm GeV} \times \left(\frac{m_\phi}{10^{15}\,{\rm GeV}}\right)^{3/2}\ ,
\end{eqnarray}
when the inflation decays into $H_u$ and $H_d$ Higgs fields through a dimension five operator,
$K\sim (S+S^\dagger) H_u H_d$.
When, instead, the inflaton decays through dimension six operators, the reheating temperature is roughly
\begin{eqnarray}
\label{eq:TRH2}
T_{R,{\rm dim} 6} \sim 10^{8-9}{\rm GeV} \times \left(\frac{m_\phi}{10^{15}\,{\rm GeV}}\right)^{5/2}\ ,
\end{eqnarray}
where we have assumed that the coefficients of the operators responsible for the decay of
the inflation are of $O(1)$.
In Eqs.~\eqref{eq:TRH1} and \eqref{eq:TRH2}, we have worked with an inflaton mass of
$m_{\phi} = O(10^{15})$\,GeV, which turns out to be a typical value (cf.\ below).%
\footnote{Even if the mass of the inflaton is as large as $10^{15}$ GeV,
such that its decay products have extremely large momenta, the inflaton
decay products thermalize soon after their production~\cite{Harigaya:2013vwa}.}
As a result, we find that high reheating temperatures can be realized rather easily,
which is quite favorable for successful leptogenesis\,\cite{Fukugita:1986hr}.

\begin{center}
\begin{figure}[t]
\includegraphics[width=\linewidth]{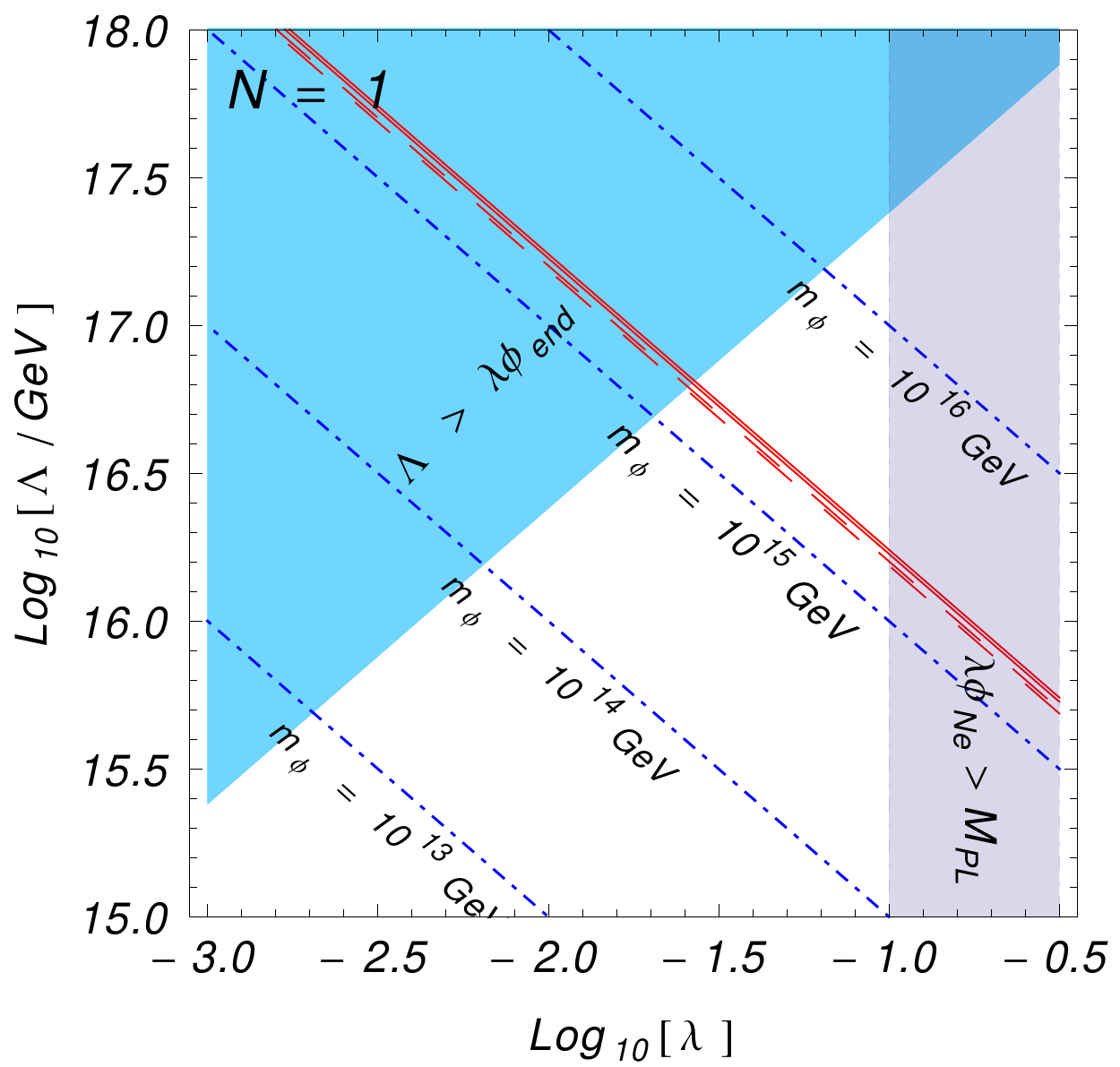}
\includegraphics[width=\linewidth]{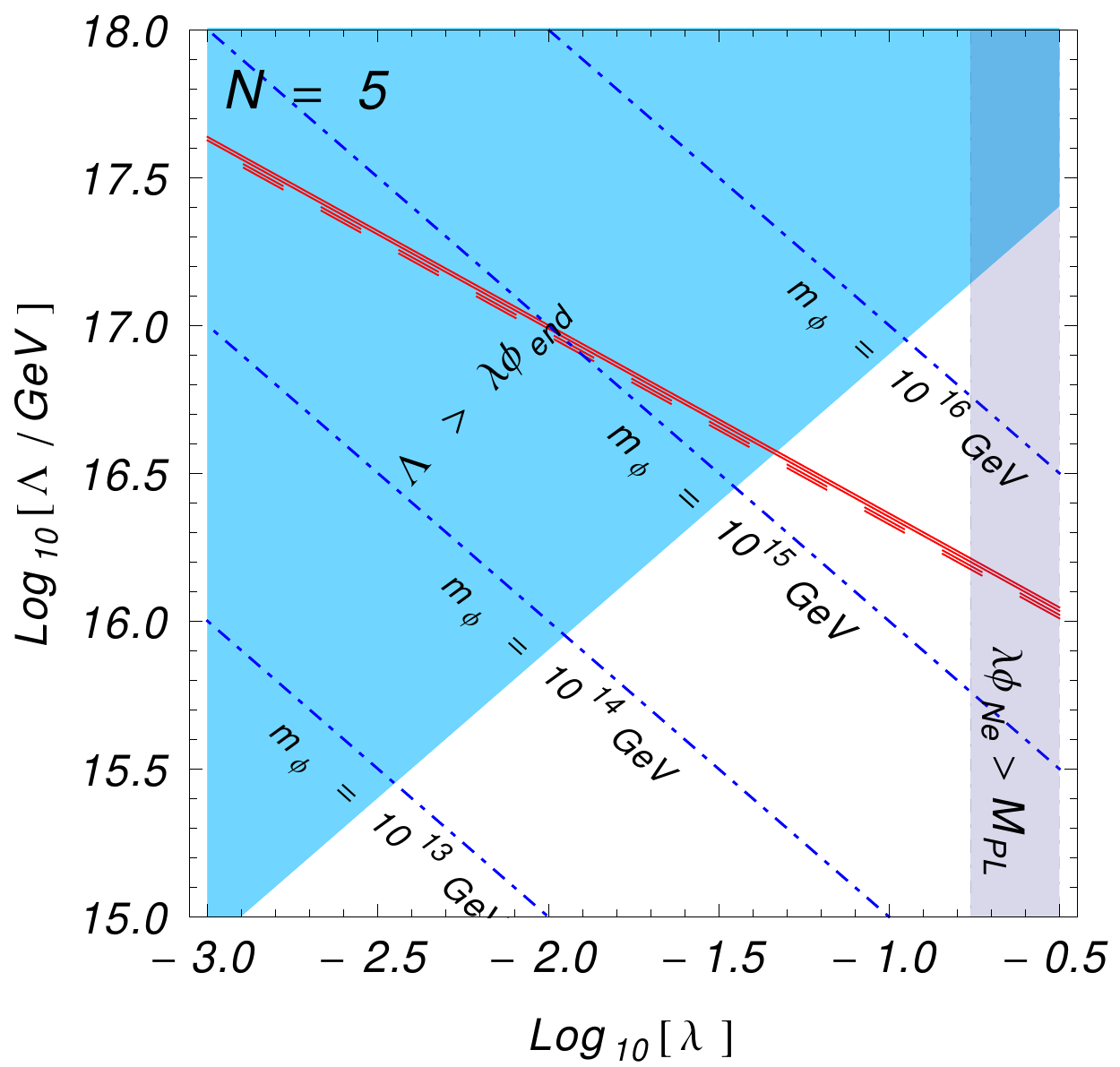}
\caption{\sl
$(\lambda,\Lambda)$ plane for $N=1$ (\textbf{upper panel}) and $N=5$ (\textbf{lower panel}).
The red lines indicate where the curvature power spectrum is
consistent with the observed value within $2\,\sigma$.
Solid and dashed lines correspond to $N_e=50$ and $60$, respectively.
We also show contour lines for the inflaton mass as blue dot-dashed lines.
The shaded regions are theoretically inaccessible because either the dynamical scale is
too large, i.e.\ $\Lambda > \l \phi_{\rm end}$,
or the coupling constant $\lambda$ is too large, i.e.\ $\lambda \phi_{Ne} > M_{\rm Pl}$,
as denoted in the figure.}
\label{fig:power}
\end{figure}
\end{center}

Now, let us discuss the predictions of our fractional power-law inflaton potential for the CMB observables 
(cf.\ also\,\cite{Alabidi:2010sf}).
Given the potential in Eq.\,(\ref{eq:spot2}),
one finds for the power spectrum $\cal{P}_{\zeta}$ of the
curvature perturbations $\zeta$\,\cite{Lyth:1998xn}
\begin{eqnarray}
{\cal P}_{\zeta}= \frac{1}{6\pi^2p^3}
\left(\frac{\Lambda}{M_{\rm Pl}}\right)^{4-p}\left(\lambda^2 pN_e\right)^{1+p/2} \ ,
\end{eqnarray}
where $N_e$ is the number of $e$-foldings before the end of inflation
when the CMB scales leave the Hubble horizon.
In Fig.\,\ref{fig:power},  the red lines mark the region in the
$(\lambda, \L)$ parameter space which is consistent with the observed curvature power spectrum,
$\ln \left(10^{10}\,{\cal P}_\zeta\right) = 3.080\pm 0.025$\,\cite{Spergel:2013rxa}
for $N=1$ and $N=5$, respectively.
The blue dot-dashed lines represent contour lines indicating the values of
the inflaton mass.
In the blue-shaded regions, the dynamical scale is rather large,
so that is also important during inflation, i.e.\ $\Lambda > \l \phi_{\rm end}$,
where $\phi_{\rm end} \simeq pM_{PL}$.
In this situation, we loose control over the inflaton potential, as it
becomes distorted by incalculable strong coupling effects.
On the other hand, in the gray-shaded regions, the coupling
$\lambda$ is too large, such that the effective mass of the heavy $Q$'s exceeds
$ M_{\rm Pl}$ during  inflation, i.e.\
$\lambda \phi_{N_e} > M_{\rm Pl}$ with $\phi_{N_e} \simeq (2pN_e)^{1/2}M_{\rm Pl}$.
In conclusion,  Fig.\,\ref{fig:power} illustrates that the observed curvature
power spectrum can be reproduced for $\lambda \simeq 10^{-(2-1)}$ and
$\Lambda \simeq 10^{16}$\,GeV, where the inflaton mass is typically of $O(10^{15})$\,GeV.

\begin{center}
\begin{figure}[t]
\includegraphics[width=\linewidth]{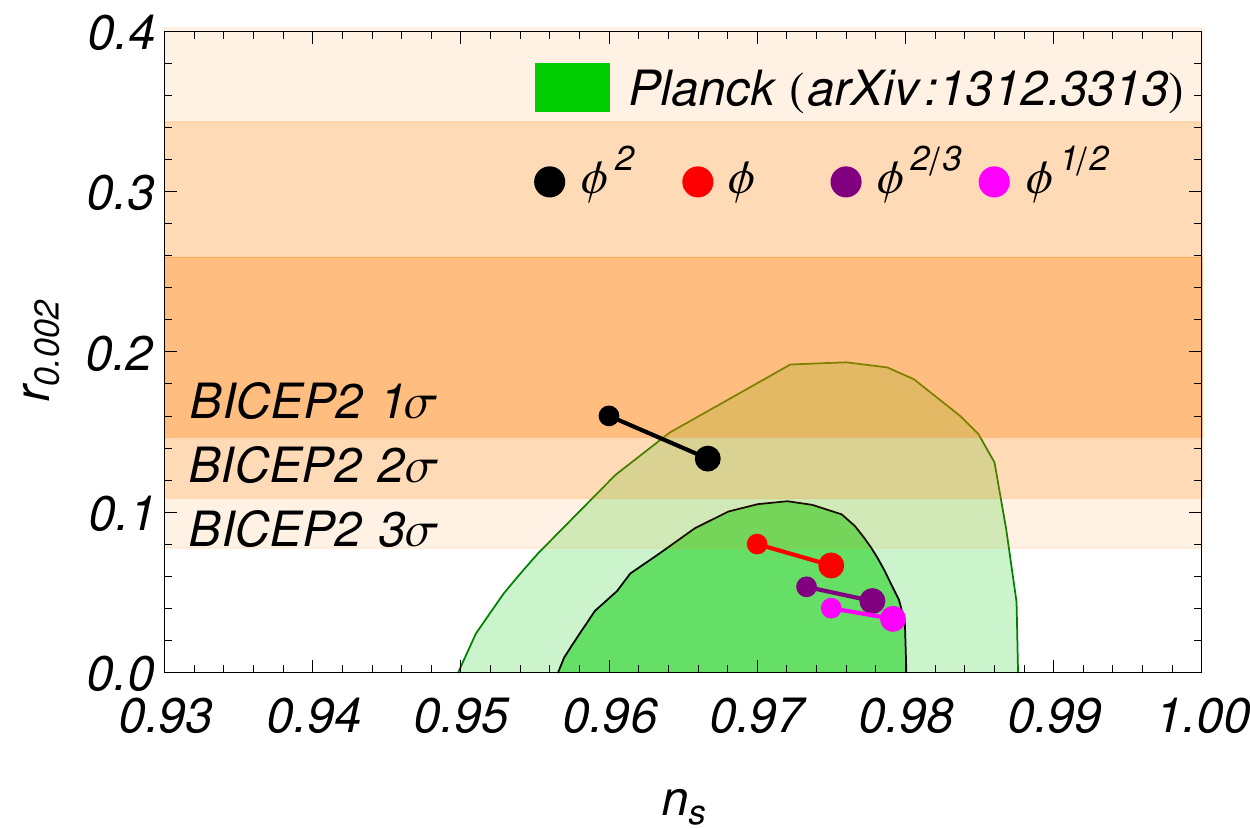}

\vspace{0.5cm}
\includegraphics[width=\linewidth]{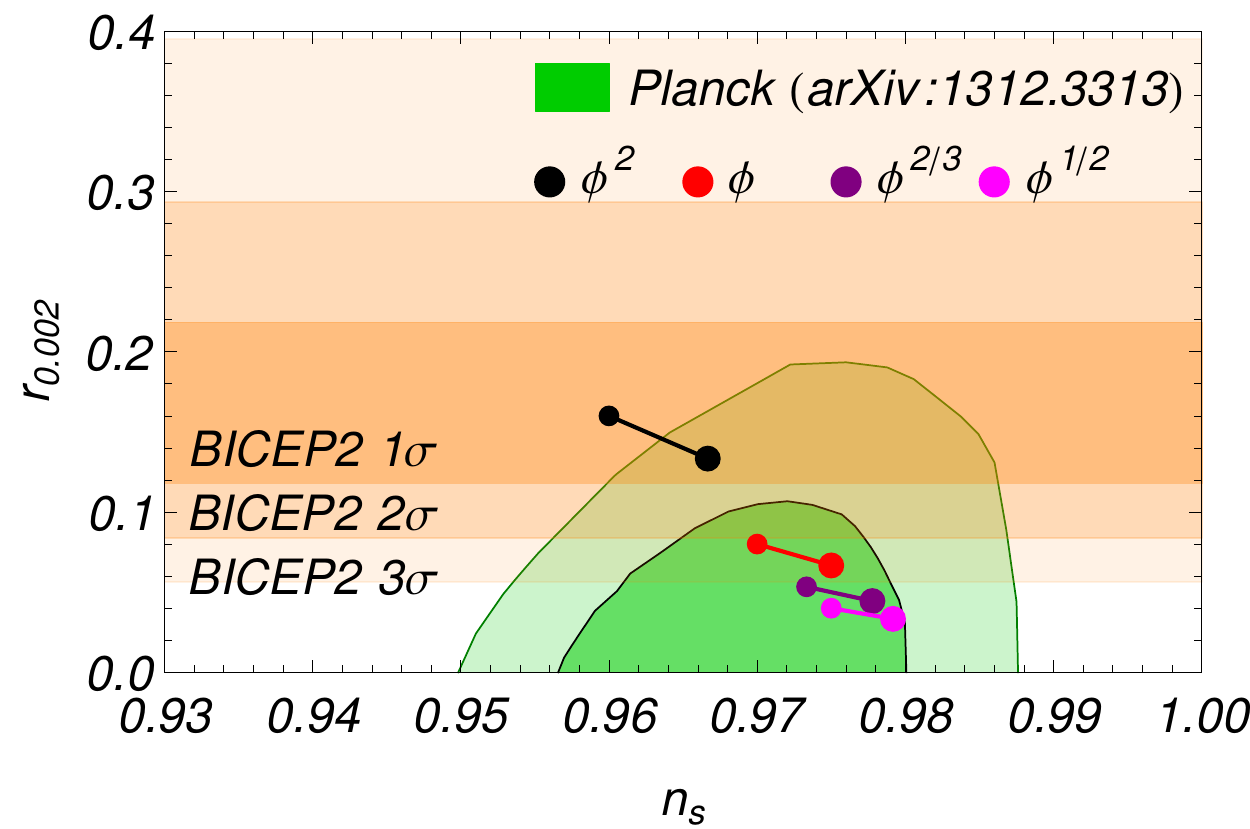}
\caption{\sl
Predicted values for $n_s$ and $r$
for $p=1$ ($N=1$), $p=2/3$ ($N=2$), and $p=1/2$ ($N=3$).
Here, $r_{0.002}$ denotes the tensor-to-scalar ratio at the pivot scale $k=0.002$\,Mpc$^{-1}$.
The predictions according to chaotic inflation with a quadratic potential are
also shown for comparison.
Small and big dots stand for $N_e=50$ and $60$, respectively.
The green contours are the constraints extracted from\,\cite{Spergel:2013rxa}.
The orange bands correspond to the $1$, $2$ and $3\,\sigma$ ranges for $r$ measured by BICEP2.
Here, the bands in the \textbf{upper panel} are purely based on the BICEP2 maps, while
the bands in the \textbf{lower panel} follow after subtracting the cross correlation
spectrum for the DDM2 foreground polarization model~\cite{BICEP2} from the raw data.}
\label{fig:nsr}
\end{figure}
\end{center}

\begin{center}
\begin{figure}[t]
\includegraphics[width=\linewidth]{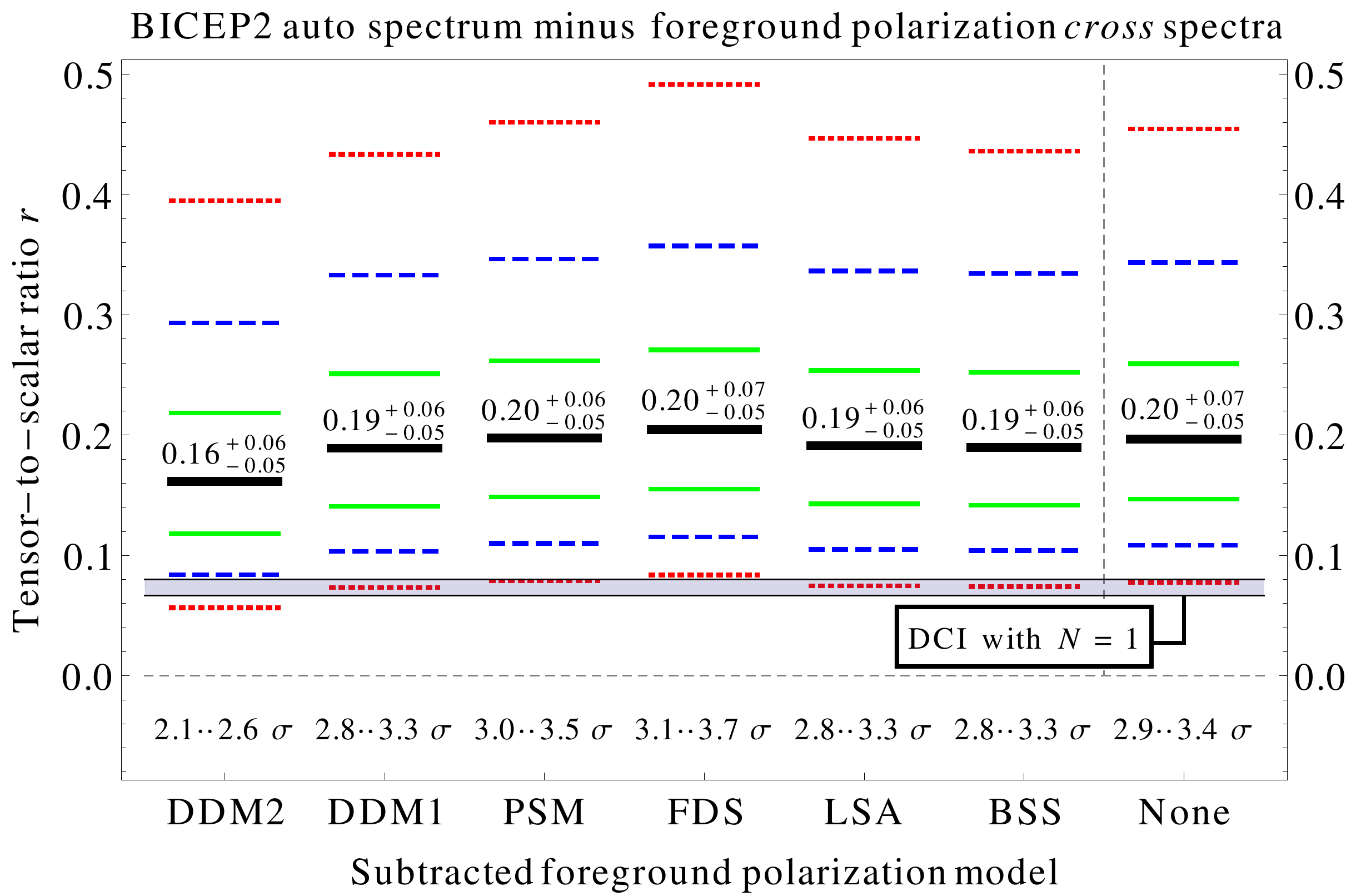}

\vspace{0.2cm}
\includegraphics[width=\linewidth]{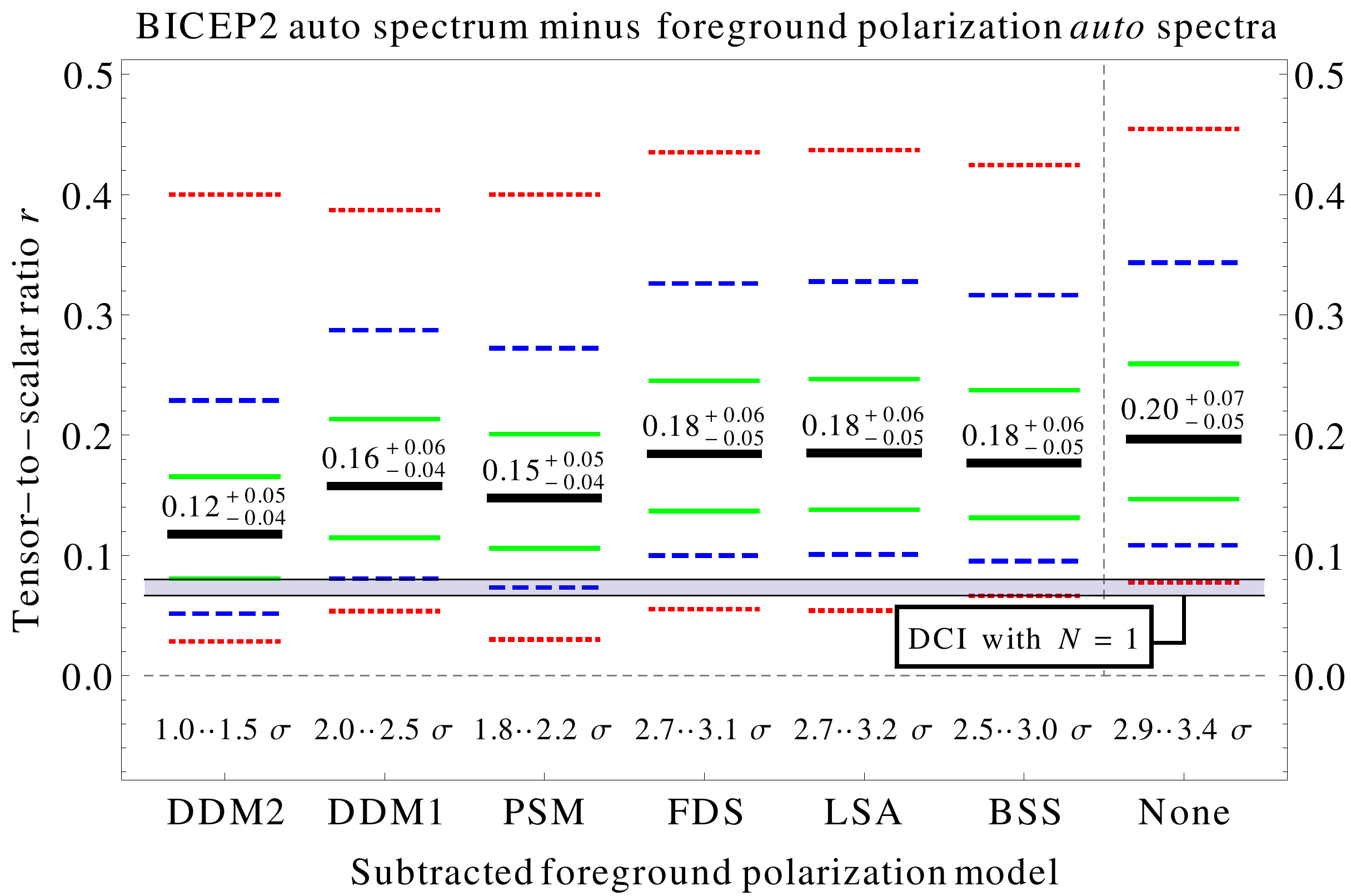}
\caption{\sl
Comparison between our prediction for $r$ in the case of dynamical chaotic
inflation (DCI) based on $SU(2)$ gauge dynamics, $r = 4/N_e \simeq 0.07\cdots0.08$,
(gray band) and the maximum likelihood values for $r$ deduced from the BICEP2
measurement and various foreground polarization models~\cite{BICEP2}.
The \textbf{upper} and the \textbf{lower panel} are respectively based on
the subtraction of the cross and auto spectra from the raw data.
For each model and subtraction procedure, we state the maximum likelihood
value for $r$, its $1\,\sigma$ range as well as the deviation of our prediction
from this value.
Here, the uncertainty in the latter figure stems from the uncertainty in $N_e$.
In addition to that, the colorful bars mark the respective $1\,\sigma$ (solid, green),
$2\,\sigma$ (dashed, blue) and $3\,\sigma$ (dotted, red) ranges for $r$.}
\label{fig:foreground}
\end{figure}
\end{center}

The spectral index $n_s$, the running of the spectral index $dn_s/d\ln k$, 
and  the tensor-to-scalar ratio $r$ of the curvature perturbations are predicted to be,
\begin{eqnarray}
 n_s = 1-\frac{p+2}{2N_e}, \quad
 \frac{d n_s}{d\ln k} = - \frac{2+p}{2N_e^2}\ ,
  \quad r = \frac{4p}{N_e} \ .
\end{eqnarray}
In the two panels of Fig.\,\ref{fig:nsr}, we show the predicted
values for $n_s$ and $r$ for $p=1$ ($N=1$), $p=2/3$ ($N=2$), and $p=1/2$ ($N=3$).
At the same time, dynamical chaotic inflation predicts
the running of the spectral index to be negligibly small.
In Fig.\,\ref{fig:nsr}, we also reproduce the constraints on $n_s$ and $r$
presented in\,\cite{Spergel:2013rxa}, in which the Planck data has been
re-analyzed taking particular care of possible systematics in the
$217\,\textrm{GHz}$ temperature map.
Furthermore, we include constraints on the tensor-to-scalar ratio
as deduced from the BICEP2 measurement.
In the upper panel of Fig.\,\ref{fig:nsr}, we indicate the 
$r$ value derived from the pure BICEP2 signal, $r = 0.20^{+0.07}_{-0.05}$,
while in the lower panel of this figure, we display the allowed $r$ range
obtained by the BICEP2 collaboration after subtracting
the arguably best model for foreground  dust polarization
from the raw data, $r = 0.16^{+0.06}_{-0.05}$.
In summary, this figure shows that models with $N>1$ are excluded
by the BICEP2 results at the $3\,\sigma$ level.
By contrast, the simplest case, i.e.\ $N=1$, is consistent
with the BICEP2 result within $3\,\sigma$.
In fact, for $N=1$ and $N_e  = 50$, dynamical chaotic inflation predicts
$r \simeq 0.08$, which deviates from the BICEP2 maximum likelihood value,
$r = 0.20$, by $2.9\,\sigma$ as well as from the corresponding value
after subtraction of the DDM2 dust foreground model, $r = 0.16$, by $2.1\,\sigma$.
A complete comparison of our prediction for $r$ in the simplest case of
an $SU(2)$ gauge group with the allowed $r$ ranges obtained for all of the various
foreground models considered by the BICEP2 collaboration
can be found in Fig~\ref{fig:foreground}.
More general scenarios of dynamical chaotic inflation, also featuring
fractional powers $p>1$, as well as their performance in view of the BICEP2
results will be addressed in \cite{Harigaya:2014zz}.

In view of the above stated deviations between our prediction of $r \simeq 0.08$
and the experimental values, it is important to note that at present there
is a tension between the constraints deduced from the Planck data and
the value measured by BICEP2, where the Planck data favors a smaller value 
of the tensor-to-scalar ratio.
Moreover, as far as our theoretical prediction is concerned, we also remark
that, if the shift symmetry is broken not only in the superpotential
but also in the K\"ahler potential, the prediction for $r$ can be
still be raised to larger
values~\cite{Kallosh:2010ug,Kallosh:2010xz,Li:2013nfa,Harigaya:2014qza}.
The same applies to generalized dynamical chaotic inflation
featuring fractional powers $p>1$~\cite{Harigaya:2014zz}.
Therefore, it is certainly premature to declare dynamical chaotic inflation
ruled out by the data at this point, in particular, the model with the simplest
gauge group $SU(2)$.
Quite the contrary, as further measurements of the CMB B-mode polarization are
being performed, dynamical chaotic inflation based on strong $SU(2)$ dynamics
might eventually develop into one of the most promising models correctly describing the data.

Another important key feature of dynamical chaotic inflation is that 
it predicts a slightly larger value for the spectral index compared with
the simplest chaotic  inflation model.
Therefore, by further observational investigation of $n_s$ and $r$,
dynamical chaotic inflation can be distinguished from the simplest model
of chaotic inflation based on a quadratic potential.%
\footnote{Fig.\,\ref{fig:nsr} slightly suggests that a smaller number of $e$-foldings
$N_e$ (lower $T_R$) is preferred in the case of the dynamical model, so as
to raise the tensor-to-scalar ratio towards the BICEP2 best-fit value,
while a larger number of $e$-foldings $N_e$ (higher $T_R$) is preferred in the case of
the quadratic potential model, so as to make the spectral index larger.}

\section{Conclusions}
In this paper, we revisited the class of models of chaotic inflation the
potential of which is generated by the dynamics of a strongly coupled supersymmetric gauge theory.
A prominent feature of this scenario of \textit{dynamical chaotic inflation}
is that the inflaton potential features a fractional power.
Contrasting dynamical chaotic inflation with the tensor-to-scalar ratio
recently observed by the BICEP2 experiment, we find that 
models with non-minimal gauge group seem to be disfavored, while
the model based on the simplest gauge group, $SU(2)$, is barely consistent
with observations.
However, since there is a tension between the Planck constraints
and the BICEP2 measurement, we need to wait for further confirmation/refutation
by other observations such as Planck, ACTpole, SPT, and POLARBEAR.
Only with additional data at hand, it will be become clear whether 
dynamical chaotic inflation is excluded or in fact a good description
of the CMB data.
Likewise, improved constraints on $n_s$ will also help to 
distinguish the dynamical chaotic inflation model from chaotic inflation based on a
quadratic potential.


\subparagraph*{Acknowledgements}
This work is supported by Grant-in-Aid for Scientific Research from the
Ministry of Education, Science, Sports, and Culture (MEXT), Japan,
No.\ 22244021 (T.T.Y.), No.\ 24740151 (M.I.), and by the World Premier
International Research Center Initiative, MEXT, Japan.
The work of K.H.\ is supported in part by a JSPS Research Fellowship
for Young Scientists.

\end{document}